\def\b{\begin{equation}}
\def\e{\end{equation}}
\def\br{\begin{eqnarray}}
\def\er{\end{eqnarray}}
\def\l{\left}
\def\r{\right}
\def\s{\scriptstyle}

\documentstyle[eqsecnum,prd,aps]{revtex}

\begin{document}

\title{A novel approach to particle production in an uniform electric field} 
\author{K.~Srinivasan\thanks{Electronic address:~srini@iucaa.ernet.in},  
T.~Padmanabhan\thanks{Electronic address:~paddy@iucaa.ernet.in}}
\address{IUCAA, Post Bag 4, Ganeshkhind, Pune 411 007, INDIA.}

\maketitle

\begin{abstract}
We outline a different method of describing scalar field particle 
production in a uniform electric field. 
In the standard approach, the (analytically continued) harmonic 
oscillator paradigm is important in describing particle production.  
In the gauges normally considered, in which the four vector potential 
depends only on the time or space coordinate, the system reduces to a 
non-relativistic effective Schr\"odinger equation with an inverted 
oscillator potential. 
The Bogolubov coefficients are determined by tunnelling in this potential. 
In the Schwinger proper time method of determining the effective 
Lagrangian, the analytically continued propagator for the usual 
oscillator system is regarded as the correct propagator for the 
inverted oscillator system and is used to obtain the gauge invariant 
result. 
\par
However, there is another gauge in which the particle production process 
has striking similarities with the one used to describe Hawking 
radiation in black holes.   
The gauge we use to describe the electric field in is the 
{\it lightcone} gauge, so named because the mode functions for a 
scalar field are found to be singular on the lightcone. 
We use these modes in evaluating the effective Lagrangian using the 
proper time technique. The key feature of this analysis is that these 
modes can be explicitly ``normalized'' by using the criterion that 
they reduce to the usual flat space modes in the limit of the 
electric field tending to zero. 
This normalization procedure allows one to determine the Schwinger 
proper time kernel without using the analytical continuation of the 
harmonic oscillator kernel that is resorted to in the standard analysis. 
We find that the proper time kernel is not the same as the analytically
continued oscillator kernel though the effective Lagrangian is the 
standard result as it should be. 
\par
We also consider an example of a confined electric field system using 
the lightcone gauge modes that has several features of interest. 
In particular, our analysis indicates that the Bogolubov coefficients, 
in taking the limit to the uniform electric field case, are multiplied 
by energy dependent boundary factors that have not been taken into 
account before. 
\end{abstract}

\draft
\pacs{PACS numbers:}

\section{Introduction}
We present a different derivation of particle production in an 
uniform background electric field in Minkowski spacetime. 
The problem addressed is that of a scalar field propagating in 
flat spacetime in such a background. 
The backreaction on the electric field due to particle production 
is not discussed but only the mechanism by which particles are 
produced is considered. 
The difference between the method described here and the standard 
analysis discussed in Refs.~\cite{schwinger,brout95,paddy91,sriram97,paddyaspects} 
arises in the gauge used. 
The electric field here is described using the {\it lightcone} 
gauge which has already been introduced in Ref.~\cite{kt99}. 
In this gauge, the mode functions are combinations of elementary 
functions and they are singular on the lightcone. 
The latter property is similar to the mode functions in  spacetimes 
with a horizon like the Schwarzchild or Rindler spacetimes with 
the modes  being singular on the horizon. 
This property of the lightcone gauge modes explicitly shows that 
particle production occurs in a similar fashion in both systems 
(though not exactly in the same way as will be subsequently shown). 
Note that these modes describe the same system as the parabolic 
cylinder functions do in the time and space dependent gauges 
used normally. 
The singularity present in the modes of the lightcone gauge manifests 
itself as the singular inverted oscillator potential in the 
other two gauges. 
These modes have the property that they can be ``normalized'' 
by a suitable physical criterion which allows one to calculate 
the Schwinger proper time kernel in a straightforward manner 
using an appropriate extension of the Feynman-Kac formula. 
Such a  normalizability property circumvents the need to regard the 
analytically continued harmonic oscillator kernel as the correct 
propagator for the inverted oscillator kernel. 
This is required in the standard analysis because the parabolic 
cylinder functions cannot be normalized in a simple manner. 
The proper time kernel determined from these modes is found to be 
different from the analytically continued harmonic oscillator kernel 
though the effective Lagrangian is the same as the standard result. 
\par
We also consider a special, spatially confined electric field 
system that can be conveniently described using the lightcone gauge. 
The system is constructed such that, in the limit of a parameter 
tending to infinity, it tends to a uniform electric field system. 
This provides us with another example of a limiting process by 
which a uniform electric field system can be described. 
The highlight of this analysis is that hitherto unaccounted for 
normalization constants appear in the expression for the 
Bogolubov coefficients. 
This occurs because, in the standard analysis, boundary conditions 
at infinity could not be explicitly considered. 
If a continuous vector potential is used to describe a confined 
electric field that varies from zero in the infinite past to a 
finite constant value in the infinite future (in the time 
dependent gauge, say), the normalization constants for positive 
frequency modes are different at these two asymptotic regions. 
Therefore, when computing the Bogolubov coefficients, this difference 
in definition of positive frequency modes appears as a 
multiplicative factor to the standard values. 
This factor is found to be the same for both the Bogolubov 
coefficients. 
As a consequence, the effective Lagrangian is found to 
match the standard Schwinger result.  
\par
This paper is organized as follows. In section~(\ref{sec:lightcone}), 
we briefly introduce the lightcone gauge. 
Section~(\ref{sec:electricefflag}) discusses the construction of 
the effective Lagrangian in the lightcone gauge.  
We show how this lightcone structure of the modes determines 
particle production. 
Then, in section~(\ref{sec:modematching}),  we discuss the confined 
electric field system. 
This system also illustrates the role of the lightcone structure 
in particle production and is reducible to the uniform electric 
field case by a well defined limiting procedure.  
We calculate the Bogolubov coefficients and effective Lagrangian 
and then take the limit to the uniform electric field case. 
Finally, in section~(\ref{sec:electricdiscussion}), we summarize 
the results of this paper.

\section{The lightcone gauge}
\label{sec:lightcone}
The lightcone gauge has already been introduced and described 
in Ref.~\cite{kt99}. 
There, we showed how the tunnelling interpretation was used 
to recover the Bogolubov coefficients. 
However, for completeness, we briefly describe the gauge here.
\par
We describe a spatially and temporally uniform electric field 
pointing along the $\hat{\bf x}$ direction by the gauge 
\b
A^{k} = {E_0\over 2}( t -x,\, x - t,\, 0,\, 0) 
\label{eqn:lightconegauge}, 
\e
where $E_0$ is the magnitude of the electric field.  
The differential equation satisfied by a massive scalar field 
$\Phi$ in an electromagnetic field background in Minkowski 
spacetime is
\b
\l[\l(\partial_k + iq A_k\r)\l(\partial^k + iqA^k\r) + m^2 \r]\Phi = 0,
\label{phi}
\e
where $m$ is the mass, $q$ is the charge and $A^k$ is the 
four potential of the electromagentic field. 
Defining the null variables
\begin{equation}
u = t - x \qquad ; \qquad v =t + x 
\label{eqn:electric41}
\end{equation}
and setting 
\b
\Phi = e^{i k_y y + ik_zz}\, e^{-i\gamma v}\, \phi(u) ,
\e
we obtain the following {\it first order} differential 
equation
\begin{equation}
2i(qE_0u - 2\gamma){d\phi \over du}  +  
(m^2 + {\bf k}_{\perp}^2 + iqE_0) \phi (u)  =  0.
\label{eqn:electric46}
\end{equation} 
The solution to the above equation is easily seen to be
\begin{equation}
\phi (u) = \left( \sqrt{qE_0\over 2}\, u - 
\gamma \sqrt{2\over qE_0}\, \right)^{i\lambda/2 \, - \, 1/2} 
\label{eqn:electric47},
\end{equation}
where
\b
\lambda = {m^2 + {\bf k}_{\perp}^2 \over qE_0} .
\label{lambda}
\e
Note that the solution for $\phi$ is an elementary function of 
the variable $u$. 
This is unlike the modes in the time or space dependent gauge 
which contain transcendental parabolic cylinder functions.  
This is the simplest mode function possible for any gauge of 
the electric field.   
This solution is singular on the null surface $t-x = \gamma/qE_0$ 
and it is for this reason that the gauge in Eq.~(\ref{eqn:lightconegauge}) 
is referred to as the lightcone gauge.
The Bogolubov coefficients can be easily calculated by constructing  
a tunnelling scenario for $\phi(u)$, calculating the transmission 
and reflection coefficients using the method of complex paths  
and then using the tunnelling interpretation to suitably interpret 
these coefficients. 
This has been done in detail in Ref.~\cite{kt99} and will 
not be repeated here. 
It was shown there that the standard result is obtained. 
\par
From the form of the lightcone gauge given in 
Eq.~(\ref{eqn:lightconegauge}), it is clear that there 
is another equivalent gauge that also gives the simplest 
mode functions possible. This gauge is of the form
\b
A^{i} = -{E_0\over 2}\l( t+x,\, t+x,\, 0,\, 0\r) .
\label{eqn:lightconegauge1} 
\e
This gauge, however, will not be considered here separately 
since its properties are very similar to that of the light 
cone gauge.
\par
In the next section, we calculate the effective Lagrangian 
using Schwinger's proper time approach. 

\section{Effective Lagrangian for Electric Field}
\label{sec:electricefflag}
The uniform electric field problem, in the time dependent gauge, 
can be essentially reduced to an effective Schr\"odinger problem 
with an inverted harmonic oscillator potential with mode functions 
that are transcendental parabolic cylinder functions. 
This effective quantum mechanical system has no ground state.   
But the basic formalism of the effective Lagrangian method 
requires that the system be in the vacuum state in the asymptotic 
past and future. It makes the implicit assumption that the electric 
field tends to zero in the asymptotic limits.  
This issue is resolved, in the path integral technique, 
by analytically continuining the simple harmonic oscillator 
kernel to imaginary frequencies~\cite{paddyaspects}. 
This analytically continued kernel is  assumed to be the 
correct kernel for the electric field system. 
Such a continuation also implies the boundary condition that 
the electric field tends to zero asymptotically.     
\par
In this section, we propose an alternative derivation of the 
effective Lagrangian result without using the harmonic 
oscillator kernel. 
The gauge we work in is the light cone gauge discussed 
in the previous section. 
The mode functions in this gauge are combinations of elementary 
functions and are singular on the light cone. 
This  singular behaviour implies that the proper time mode functions 
cannot be normalized by the usual Schr\"odinger normalization 
condition. 
(This non-normalizability of the proper time modes also occurs 
in the standard approach because of the presence of the parabolic 
cylinder functions which do not have the required asymptotic 
behaviour.) 
However, this can be circumvented by demanding that these modes 
reduce to the  usual flat space modes in the limit of the electric 
field $E_0 \to 0$. 
Imposing this normalization condition is equivalent to the 
analytic continuation that is resorted to in the standard 
approach and gives the correct result. 
The lightcone structure of these modes plays an important 
role in determining particle production as will be shown. 
The entire analysis is conveniently done in the $(u,v,y,z)$ 
coordinate system where $u=t-x$ and $v=t+x$ are the usual 
null coordinates.     
\par
This section is divided into two parts. In the first part, 
the proper time kernel is calculated for the case $E_0 = 0$ 
in the $(u,v,y,z)$ coordinate system  in order to motivate 
the discussion for the case $E_0 \neq 0$ which will be 
considered in the next part. The effective Lagrangian will 
be calculated subsequently. 
 
\subsection{Proper time kernel for $E_0 = 0$}
\label{subsec:Minkowskikernel}
For the case $E_0 = 0$, the proper time effective 
Schr\"odinger equation in the $(u,v,y,z)$ coordinate 
system for a scalar field of mass $m$ is 
\b
\l( 4\partial_u\partial_v - \nabla_{\perp}^2 + m^2\r)\Phi = E\Phi,
\label{minkkernel1}
\e
where $E$ is the ``energy'' corresponding to the proper time $s$. 
The solution to the above equation is of the form
\b
\Phi = N e^{i{\bf k}_{\perp}\cdot {\bf r}_{\perp}}e^{-i\alpha u}
e^{-i\gamma v},
\label{minkkernel2}
\e
where $N$ is a normalisation constant to be determined and 
$\alpha$, $\gamma$ and ${\bf k}_{\perp} = (0, k_y, k_z)$ are 
arbitrary constants taking values in the range $(-\infty, \infty)$ 
such that 
\b
E = m^2 + {\bf k}_{\perp}^2 - 4\alpha\gamma .
\label{minkkernel3}
\e
We normalize $\Phi$ by the usual Schr\"odinger prescription to 
obtain the normalized wavefunctions
\b
\Phi = {\sqrt{2} \over (2\pi)^2}e^{i{\bf k}_{\perp}\cdot 
{\bf r}_{\perp}}e^{-i\alpha u}e^{-i\gamma v}.
\label{minkkernel4}
\e
The proper time kernel, using the Feynman-Kac expression, is
\br
K(a,b;s) &=& {2 \over (2\pi)^4}\int\! d^2{\bf k}_{\perp}d\alpha d\gamma 
e^{i{\bf k}_{\perp}\cdot ({\bf a}_{\perp}-{\bf b}_{\perp})} 
e^{-i\alpha (u_a - u_b)} e^{-i\gamma(v_a-v_b)} e^{-iEs} \nonumber \\
&=& {2\pi \over (2\pi)^4is} e^{i({\bf a}_{\perp}^2 - {\bf b}_{\perp}^2)/4s}
e^{-im^2s} (2\pi)\int\!d\gamma \, 
\delta\l[4\gamma s - (u_a - u_b)\r] e^{-i\gamma(v_a - v_b)} 
\nonumber \\
&=& {1 \over 16\pi^2 is^2} e^{-im^2s}e^{i({\bf a}_{\perp}^2 - 
{\bf b}_{\perp}^2)/4s} e^{-i(u_a-u_b)(v_a-v_b)/4s} 
\nonumber \\
&=& {1 \over 16\pi^2 is^2} e^{-im^2s} e^{-i(a-b)^2/4s}
\label{minkkernel5},
\er
where we have substituted for $E$ from Eq.~(\ref{minkkernel3}), 
$a = (u_a, v_a, a_y, a_z)$, $b = (u_b, v_b, b_y, b_z)$, 
${\bf a}_{\perp} = (0, a_y,a_z)$, ${\bf b}_{\perp} = (0, b_y,b_z)$ 
and $\delta(x)$ is the one dimensional Dirac delta function. 
The above result is seen to be the standard result for 
a complex scalar field. 

\subsection{Proper time kernel for $E_0 \neq 0$}
\label{subsec:electrickernel}
Now, we are ready to consider the case of a uniform electric 
field pointing along the $\hat{\bf x}$ direction with a 
magnitude $E_0$. The gauge we consider is the light cone gauge.
The proper time effective Sch\"odinger equation in this case is 
\b
\l[ \l(\partial_i + iqA_i\r)\l(\partial^i + iqA^i\r) + m^2\r]\Psi 
= E\Psi .
\label{electrickernel1}
\e
The lightcone gauge, in the $(u,v,y,z)$ coordinate system, is
\b
A^i = E_0\l(u,\, 0,\, 0,\, 0\r)
\label{electrickernel2}.
\e
The solutions to Eq.~(\ref{electrickernel1}) are given by
\b
\Psi = N e^{i{\bf k}_{\perp}\cdot {\bf r}_{\perp}}e^{-i\gamma v} 
\l[ \sqrt{qE_0 \over 2} u - 
\sqrt{2 \over qE_0}\gamma\r]^{i\varrho -{1\over 2}} ,
\label{electrickernel3}
\e
where $N$ is a normalisation constant to be determined, 
$\gamma$, $\varrho$ and ${\bf k}_{\perp}$ are arbitrary 
constants and the energy $E$ is given by the relation 
\b
E = m^2 + {\bf k}_{\perp}^2 - 2qE_0\varrho \; .
\label{electrickernel4}
\e
Since the above mode functions are singular on the surface 
$u = 2\gamma/qE_0$, it can easily be shown that they cannot be 
normalized by the usual Schr\"odinger prescription. 
To make progress, we impose the condition that, in the limit 
of $qE_0 \to 0$, $\Psi$ reduces to the usual Minkowski mode 
functions given in (\ref{minkkernel4}). 
This implies that the ``normalized'' mode functions must be of 
the form
\b
\Psi = {\sqrt{2} \over (2\pi)^2} 
\l(\sqrt{2 \over qE_0}\gamma\r)^{-{\s 2i\alpha \gamma \over \s qE_0} 
+ {\s 1\over \s 2}} e^{i{\bf k}_{\perp}\cdot {\bf r}_{\perp}}e^{-i\gamma v} 
\l[ \sqrt{2 \over qE_0}\gamma - 
\sqrt{qE_0 \over 2} u \r]^{{\s 2i\alpha\gamma \over \s qE_0} -
{\s 1\over \s 2}},
\label{electrickernel5}
\e
where a new constant $\alpha$ has been defined such that 
\b
E = m^2 + {\bf k}_{\perp}^2 - 4\alpha \gamma ,
\label{electrickernel6}
\e
just as in (\ref{minkkernel3}) with $\varrho = 2\alpha\gamma/qE_0$. 
Extending the definition of the Feynman-Kac formula for 
such a system, we have
\br
K(a,b;s) &=& {2 \over (2\pi)^4}\sqrt{2\over qE_0} \int\! 
d^2{\bf k}_{\perp}d\alpha d\gamma \,\gamma  e^{i{\bf k}_{\perp}\cdot 
({\bf a}_{\perp}- {\bf b}_{\perp})}e^{-i\gamma (v_a-v_b)} 
e^{-i(m^2 + {\bf k}_{\perp}^2 -4\alpha \gamma)s} \nonumber \\
& \times & \l[ \sqrt{2 \over qE_0}\gamma -\sqrt{qE_0 \over 2}
u_a \r]^{{ \s 2i\alpha\gamma \over \s qE_0} -{\s 1 \over \s 2}}
\l[ \sqrt{2 \over qE_0}\gamma - 
\sqrt{qE_0 \over 2}u_b \r]^{-{\s 2i\alpha\gamma \over \s qE_0} -
{\s 1\over \s 2}} 
\label{electrickernel7}.
\er
Doing the integrals over $k_y$ and $k_z$ and defining new 
dimensionless variables
\b
\alpha' = \sqrt{2\over qE_0}\alpha, \qquad \gamma' = 
\sqrt{2\over qE_0}\gamma,
\label{electrickernel8}
\e
the expression for the kernel becomes
\br
K(a,b;s) &=& {2 \over (2\pi)^3is}\l[{qE_0 \over 2}\r]^{3/2} 
e^{-im^2s}e^{i({\bf a}_{\perp} - {\bf b}_{\perp})^2/4s}
\int\! d\alpha' d\gamma' \, \gamma' e^{-i\gamma' 
\sqrt{qE_0/2}(v_a-v_b)}  \nonumber \\
&& \times  \;  e^{2iqE_0s \alpha'\gamma'} \, 
\l[\l(\gamma'-u_a\sqrt{qE_0/2}\r)\l(\gamma'-u_b
\sqrt{qE_0/2}\r)\r]^{-{1\over 2}}  \nonumber \\ 
&& \times \; \exp\l(i\alpha'\gamma'\ln\l[{\gamma'-u_a\sqrt{qE_0/2} 
\over \gamma'-u_b\sqrt{qE_0/2}}\r] \r)
\label{electrickernel9}.
\er
When $u_a \neq u_b$ and assuming further that $u_a>u_b$ for 
definiteness, the integral over $\alpha'$ gives a Dirac delta 
function which can be easily evaluated to give the result
\br
K(a,b;s) &=& {1 \over 16\pi^2 is} \, {qE_0 \over \sinh(qE_0s)} e^{-im^2s}
e^{i({\bf a}_{\perp} - {\bf b}_{\perp})^2/4s} \nonumber \\
&& \times \exp\l(-{iqE_0 \l(u_a - u_be^{-2qE_0s}\r)(v_a-v_b) 
\over 2\l(1-e^{-2qE_0s}\r) }\r) .
\label{electrickernel11}
\er
In the limit of $qE_0 \to 0$, it is easily checked that the above 
result reduces to the free field result in Eq.~(\ref{minkkernel5}). 
This result is clearly not the same as the proper time kernel of the analytically 
continued simple harmonic oscillator. 
This difference merely reflects the choice of gauge in each case.  
\par
The effective Lagrangian can be calculated using the above kernel by 
setting $v_a = v_b$, ${\bf a}_{\perp} = {\bf b}_{\perp}$ and 
taking the limit $u_a \to u_b$. This gives
\br
L_{\rm eff} &=& -i\lim_{a\to b}\int_0^{\infty} \! 
{ds \over s}K(a,b;s) \nonumber \\
&=& -{1 \over 16\pi^2} \int_0^{\infty}\! {ds \over s^2} 
e^{-im^2 s} {qE_0 \over \sinh(qE_0s)}
\label{electrickernel12},
\er
which indeed is the standard result~\cite{paddyaspects}. 
The imaginary part of the effective Lagrangian can be calculated 
in the usual way by using standard contour integration techniques. 
Note that the final answer is seen to be valid even if $u_a \neq u_b$ 
(but with $v_a = v_b$ and ${\bf a}_{\perp} = {\bf b}_{\perp}$) 
and so is perfectly well defined in the limit $u_a \to u_b$.  
Therefore, we see that the standard result is recovered in a rather 
simple and straightforward manner. The role of the light cone 
structure appears to make no difference to the production of particles as to be 
expected of a gauge invariant result. 
The following subtlety however, is worth noting. 
Let us impose the condition $u_a = u_b$ {\it before} the 
evaluation of the integral over $\alpha'$ in 
Eq.~(\ref{electrickernel9}). 
In this case, the last exponential  containing the logarithmic 
term does not contribute.  Evaluating the integral over $\alpha'$, 
one has
\br
K(a,b;s) &=& {2 \over (2\pi)^2is}\l[{qE_0 \over 2}\r]^{3/2} 
e^{-im^2s}e^{i({\bf a}_{\perp} - {\bf b}_{\perp})^2/4s}
\int\! d\gamma' \, \gamma' e^{-i\gamma' \sqrt{qE_0/2}(v_a-v_b)} 
\nonumber \\
&& \times \l(\gamma'-\sqrt{qE_0/2}u_a\r)^{-1} \, \delta(2qE_0s\gamma') 
\nonumber \\
&\equiv& 0 .
\label{electrickernel10}
\er
The kernel vanishes with the consequent result that the  effective 
Lagrangian is {\it identically zero}.  
A possible  way of understanding this sensititivity to the 
order of operations is as follows. 
The presence of the electric field produces a singularity 
on the light cone at each spacetime point (with the singularity 
at a point $x$ occuring at a time $t=x+(2\gamma/E_0)$). 
In order to have pair production, the electric field modes have 
to propagate past this lightcone singularity. 
Imposing the condition $u_a = u_b$ before the evaluation of 
the integral over $\alpha'$ in Eq.~(\ref{electrickernel9}) 
implies that this propagation across the singularity does not 
take place with the result that the kernel and the effective 
Lagrangian do not acquire an imaginary part. 
Hence no particles are produced. The zero result is primarily 
the consequence of the normalization criteria used to normalize 
the modes and just means that there is no vacuum 
polarization term present.  
When $u_a\neq u_b$, the evaluation of the Dirac Delta function 
to give the result in Eq.~(\ref{electrickernel11}) ensures 
propagation of these modes across the singularity. 
Since the contribution to the kernel from the singularity, which 
results in the appearance of an imaginary term, is independent 
of $u_a$ or $u_b$ (this is analogous to a tunnelling situation 
where the tunnelling coefficients are independent of the initial 
and final coordinates and arise only from the singularities 
and turning points present in the potential in the 
complex plane; see, for example, Ref.~\cite{kt99}), 
the final answer is well defined even in the limit of 
$u_a \to u_b$. 
We can therefore conclude that particle production in an 
uniform electric field is dependent on the light cone structure 
of the electric field modes. 
This clearly shows that electric field particle production 
is essentially a tunnelling process. 
In the time and space dependent gauges, the singular potential 
was responsible for particle production while in the light cone 
gauge, it is the singularity present on the lightcone. 
 
\section{A confined electric field system}
\label{sec:modematching}
It was mentioned earlier that the uniform electric field problem, 
in the purely time dependent gauge, can be essentially reduced 
to an effective Schr\"odinger problem with an inverted harmonic 
oscillator potential. Since this system does not possess the 
required asymptotic properties, the effective Lagrangian has 
to be calculated in a suitable fashion (in the path integral 
method, the analytic continuation of the proper time kernel 
of the simple harmonic oscillator to imaginary frequencies 
provides the solution). In order to explicitly justify the 
method used to compute the effective Lagrangian, one has to 
consider a system where the electric field is temporally bounded. 
That is, one should assume a continuous four vector potential 
that corresponds to zero electric field everywhere in the distant past 
and future. This four vector potential should also contain a 
parameter that enables this system to tend to the uniform field 
case in a suitable limit. By appropriate mode matching at the 
boundaries (or by determining the exact solution for a smoothly 
varying electric field) and calculating the Bogolubov coefficients 
and subsequently the effective Lagrangian, it ought to be possible 
to verify if the methods used in the standard calculation are 
justified by taking an appropriate limit to the uniform field case. 
\par
One such electric field which is mathematically tractable is a 
time varying homogeneous electric field system of the form
\b
{\bf E} = {E_0 \over \cosh^2(\omega t)} \hat{\bf x} 
\label{tempfield1},
\e
which tends to zero in the infinite past and future~\cite{popov72} 
(see also Refs.~\cite{nandn70,nikishov70two} for related work). 
The above example admits an exact solution for the mode functions 
in terms of hypergeometric functions (see \cite{mandf53}, 
Part 2, pp.1651-1660). 
In the limit of $\omega \to 0$, it is clear that the system 
tends to an uniform electric field system. This system can be 
reduced to an effective Schr\"odinger equation in the $t$ coordinate. 
By analysing this effective Schr\"odinger system, it can be 
shown that, in the limit $\omega \to 0$, the transmission and 
reflection coefficients tend to the standard values thus 
showing that these values are obtainable using a well defined 
limiting process. Thus, the effective Lagrangian can be obtained 
in a consistent manner which is free of the issue raised in 
the previous paragraph about the non-asymptotic behaviour of 
the uniform electric field system. 
\par 
In the example discussed above, the boundary conditions at 
termporal infinity were not imposed properly. 
Though the complete solution is given in terms of hypergeometric 
functions which have the required asymptotic behaviour, 
it should be noted that an extra normalization factor arises 
if these modes are matched to the standard Minkowski mode functions 
as mentioned in the introduction. 
This extra term modifies the expressions for the Bogolubov 
coefficients and thus the number of particle pairs created 
is different. 
However, the relative probability of pair creation, which is 
quantified by the reflection coefficient $R$ in the temporally 
varying electric field~\cite{popov72}, remains unchanged. 
This occurs because both the Bogolubov coefficients are modified 
in exactly the same way by a multiplicative term.
\par 
We shall now study an example of a confined electric field system 
that is conveniently described using the lightcone gauge. 
Recall that the modes in this gauge are singular on the lightcone 
surface. 
This is very similar to the black hole system where the modes are 
singular on the horizon which is also a null surface. 
Particle production described in such a gauge appears to be 
remarkably similar to that occuring in a black hole system  
with the presence of a null surface playing an important role 
(also see the concluding paragraphs of 
section~(\ref{sec:electricefflag})). 
This system is constructed such that, in the limit of a 
suitable parameter tending to infinity, it tends to a 
uniform electric field system. 
It also clarifies the issue raised in the previous paragraph 
by showing that the Bogolubov coefficients, in this limit, 
are modified by an extra factor that arises due to mode matching at the boundaries.
   
\par
Consider a vector potential that is  continuous in the null 
coordinate $u=t-x$ of the form
\b
A^i = \l\{ \begin{array}{lll}
(0,\, 0,\, 0,\, 0) & \qquad u \leq u_1 &  
\qquad ({\rm in \; region}) \\
E_0 (u-u_1,\, 0,\, 0,\, 0) & \qquad u_1 < u < u_2  & 
\qquad ({\rm region \; II}) \\ 
E_0(u_2-u_1,\, 0,\, 0,\, 0) & \qquad u \geq u_2 &  
\qquad ({\rm out \; region}) \\
\end{array} \r. \; ,
\label{matchingmodes1a}
\e
where $u_1$ and $u_2$ are constants. The electric field, 
charge density $\rho$ and current density ${\bf j}$ for 
this system are
\br
{\bf E} &=& E_0 \theta(u-u_1)\theta(u_2 - u)\hat{\bf x}, 
\nonumber \\
\rho &=& {E_0 \over 4\pi} \l[ \delta(u_2 - u) - \delta(u-u_1)\r] , 
\qquad {\bf j} = \l(\rho,\,0,\,0\r)
\label{matchingmodes1b},
\er 
where $\theta(x)$ the step function. 
The above electric field propagates along the null geodesic 
$u=constant$. 
At any particular point in space, the electric field switches 
on and off for a finite time interval starting from some particular 
time that is dependent on the location of this point. 
That is, the electric field, at any fixed point $x$ in space, 
switches on at $t_i = x + u_1$ and switches off at 
$t_f = x + u_2$ which therefore implies that the interval 
during which the electric field is on at any point in space 
is $t_f-t_i = u_2 - u_1 = T$. 
During this time interval, the electric field at that point 
is constant. 
The charge and current density configuration required to set 
up such a system is clearly unfeasible since it involves 
positive and negative charges moving at the speed of light 
along the null lines $u=u_2$ and $u=u_1$ respectively. 
Though such a system is physically unrealizable, it is 
neverthless worth studying since it is mathematically simple 
and tends to a uniform electric field system by a concrete 
limiting procedure.  
\par   
From the form of $A^i$ in Eq.~(\ref{matchingmodes1a}), we see 
that our calculations can be conveniently done in the 
$(u.v.y,z)$ coordinate system where $u=t-x$ and $v=t+x$. 
The advantage is that the scalar wave equation reduces to solving 
first order equations in the $u$ and $v$ variables. 
Therefore, imposing boundary conditions in the $u$ variable 
involves just matching the modes at the boundary and {\it not} 
the first derivatives. 
The scalar wave equation in the $(u,v,y,z)$ coordinate system 
with a vector potential of the form $A^i = (f(u),\, 0,\, 0,\,0)$ is
\b
\l[ 4\partial_u\partial_v + 2iqf(u)\partial_u + iqf'(u) - 
\nabla_{\perp}^2 + m^2 \r]\Psi = 0
\label{matchingmodes1},
\e
where $f'(u) = df/du$, 
$\nabla_{\perp}^2 = \partial_y^2 + \partial_z^2$ and $q$ and 
$m$ are the charge and mass of the scalar field respectively. 
\par 
The flat space modes in the ``in region'', $u<u_1$, are
\b
\Psi_{\rm in} = N_{\rm in} e^{i{\bf k}_{\perp}\cdot{\bf r}_{\perp}} 
e^{-i\gamma_{\rm in} v}e^{-i\alpha_{\rm in} u},\qquad m^2 + 
{\bf k}_{\perp}^2 = 4\alpha_{\rm in}\gamma_{\rm in} ,
\label{matchingmodes2}
\e
where $\gamma_{\rm in}$ and $\alpha_{\rm in}$ are arbitrary 
constants satisfying the relation on the right and $N_{\rm in}$ 
is a normalization constant to be determined. 
The other independent mode is $\Psi_{\rm in}^*$ which has an 
equivalent expression. 
We would now like to determine the conditions on $\alpha_{\rm in}$ 
and $\gamma_{\rm in}$ so that $\Psi_{\rm in}$ can be identified 
as a positive frequency mode. 
This can be done by noting that the normalized Minkowski positive 
frequency mode $\Phi$ in the $(t,x,y,z)$ coordinate system in 
the same region can be written in the alternative form   
\br
\Phi_{\rm in} &=& {1 \over \sqrt{(2\pi)^3 2 \omega_{\rm in}}} 
e^{i{\bf k}_{\perp}\cdot{\bf r}_{\perp}} e^{ik_x x} 
e^{-i\omega_{\rm in} t} \nonumber \\
&=& {1 \over \sqrt{(2\pi)^3 2 \omega_{\rm in}} } 
e^{i{\bf k}_{\perp}\cdot{\bf r}_{\perp}} 
e^{-i(\omega_{\rm in} -k_x)v/2} e^{-i(\omega_{\rm in} + k_x)u/2},
\label{matchingmodes3}
\er
where $\omega_{\rm in} = \sqrt{k_x^2 + {\bf k}_{\perp}^2 + m^2} > 0$. 
Comparing the modes in Eq.~(\ref{matchingmodes2}) and Eq.~(\ref{matchingmodes3}), 
one sees that, for $\Psi_{\rm in}$ 
to be a positive frequency mode, one must have 
\b
\gamma_{\rm in} = {1 \over 2}(\omega_{\rm in} -k_x) > 0, 
\quad \alpha_{\rm in} = {1 \over 2}(\omega_{\rm in}  +k_x) > 0, 
\quad N_{\rm in} = {1\over \sqrt{(2\pi)^3 2\omega_{\rm in}}},
\label{matchingmodes4}
\e
which therefore implies that  
$4\alpha_{\rm in} \gamma_{\rm in} = \omega_{\rm in}^2 - k_x^2$. 
A similar argument shows that $\Psi_{\rm in}^*$ can be 
identified with negative frequency modes. 
\par
In the previous section, a suitable normalization criterion 
for the electric field modes was introduced. 
We apply the same to the electric field modes in region 
${\rm II}$ i.e. we demand that these modes reduce to the 
standard Minkowski modes in the limit $E_0 \to 0$. 
Keeping in mind, the relations in  Eq.~(\ref{matchingmodes4}), 
the normalized electric field modes in the region $u_1<u<u_2$ 
(denoted by $\Psi_{\rm II}$) are given by
\br
\Psi_{\rm II} &=& N_{\rm in} \l(\sqrt{2 \over qE_0}
\gamma_{\rm in}\r)^{-{\s 2i\alpha_{\rm in} \gamma_{\rm in} 
\over \s qE_0} + {\s 1\over \s 2}} e^{i{\bf k}_{\perp}\cdot 
{\bf r}_{\perp}}e^{i\alpha_{\rm in}u_1} e^{-i\gamma_{\rm in} v} 
\nonumber \\
&& \qquad \times \quad \l[ \sqrt{2 \over qE_0}\gamma_{\rm in} - 
\sqrt{qE_0 \over 2} (u-u_1) \r]^{{\s 2i\alpha_{\rm in}\gamma_{\rm in} 
\over \s qE_0} -{\s 1\over \s 2}}.
\label{matchingmodes5}
\er
The other independent mode is $\Psi_{\rm II}^*$ which has an 
equivalent expression. Since the normalization criterion for 
$\Psi_{\rm II}$ has been chosen such that, in the limit of 
$E_0 \to 0$, they reduce to the standard modes in 
Eq.~(\ref{matchingmodes2}), we identify $\Psi_{\rm II}$ as 
the {\it electric field positive frequency vacuum mode}. 
Similarly, $\Psi_{\rm II}^*$ can be identified as the electric 
field negative frequency vacuum mode. 
\par
Note that $\Psi_{\rm II}$ and $\Psi_{\rm II}^*$ are singular on 
the light cone surface $u = u_s = u_1 + 2\gamma_{\rm in}/qE_0$. 
For particle production to take place, it is {\it necessary} that 
the condition $u_1 < u_s< u_2$ hold. This arises as follows: 
If the boundary condition, that for $u<u_s$, only positive 
frequency electric field modes are present, is imposed, then  
the electric field modes for $u>u_s$ are not pure positive 
frequency modes but are a combination of both positive and 
negative frequency modes because of the singularity at $u=u_s$. 
This implies particle production and is possible only if $u_s$ 
lies between $u_1$ and $u_2$. Substituting for $\gamma_{\rm in}$ 
from Eq.~(\ref{matchingmodes4}), this condition can be written as
\b
\omega_{\rm in} - k_x < qE_0T .
\label{matchingmodes6}
\e
For this electric field, only those vacuum modes with wave 
vectors $(k_x,k_y,k_z)$ satisfying the above condition are excited 
and hence contribute to particle production. 
This condition will be used when the effective Lagrangian 
is calculated.   
\par
Finally, consider the ``out region''. 
The normalized Minkowski positive frequency modes in the $(t,x,y,z)$ 
coordinate system in this region are
\b
\Phi_{\rm out} = {1 \over \sqrt{\smash (2\pi)^3\l(2\omega_{\rm out} - 
qE_0T\r)}}  e^{i{\bf k}_{\perp}\cdot{\bf r}_{\perp}} 
e^{i\bar{k}_xx} e^{-i\omega_{\rm out}t}  ,
\label{matchingmodes7}
\e
where $\omega_{\rm out}$, $\bar{k}_x$ and $(k_y,k_z)$ satisfy 
the relation
\b
\l(\omega_{\rm out} - qE_0T/2\r)^2 - \l(\bar{k}_x + qE_0T/2\r)^2 =  
{\bf k}_{\perp}^2 + m^2 .
\label{matchingmodes8}
\e
Thus, by analogy with that done for the ``in region'', the 
normalized positive frequency modes in the $(u,v,y,z)$ coordinate 
system are 
\b
\Psi_{\rm out} = N_{\rm out}  e^{i{\bf k}_{\perp}\cdot{\bf r}_{\perp}} 
e^{-i\gamma_{\rm out}v} e^{-i\alpha_{\rm out}u}  ,
\label{matchingmodes9}
\e
with the identifications
\b
\gamma_{\rm out} = {1 \over 2}(\omega_{\rm out} -\bar{k}_x), 
\quad \alpha_{\rm out} = {1 \over 2}(\omega_{\rm out}  +\bar{k}_x), 
\quad N_{\rm out} = {1\over \sqrt{\smash (2\pi)^3 \l(2\omega_{\rm out} 
- qE_0T\r)}} .
\label{matchingmodes10}
\e
The constant $\gamma_{\rm out}$, or equivalently $\bar{k}_x$, is 
determined later by using the matching conditions at the 
boundary $u=u_2$.  
The other independent mode is $\Psi_{\rm out}^*$ which has an 
equivalent expression. 
\par
Thus, in all three regions, we have
\b
\Psi = \l\{ \begin{array}{ll}
N^{\rm in}_1 \Psi_{\rm in} + N^{\rm in}_2\Psi_{\rm in}^* & 
\quad u \leq u_1 \\ 
N^{\rm II}_1 \Psi_{\rm II} + N^{\rm II}_2\Psi_{\rm in}^* & 
\quad u_1< u < u_s \\
N^{\rm II}_3 \Psi_{\rm II} + N^{\rm II}_4\Psi_{\rm in}^* & 
\quad u_s< u < u_2 \\
N^{\rm out}_1 \Psi_{\rm out} + N^{\rm out}_2\Psi_{\rm out}^* & 
\quad u \geq u_2 \\
\end{array} \r.
\label{matchingmodes11} ,
\e
where $N^{\rm in}_1$, $N^{\rm in}_2$, \ldots , $N^{\rm out}_2$ 
are constants to be determined. 
We now impose the boundary condition that, for $u<u_1$, only positive 
frequency Minkowski modes are present. 
This immediately implies that $N^{\rm in}_2 = N^{\rm II}_2=0$. 
Matching the modes at the boundary $u=u_1$ gives 
$N^{\rm II}_1 = N^{\rm in}_1$.
These positive frequency electric field modes propagate past the 
singularity at $u=u_s$ to become a combination of positive 
and negative frequency modes with amplitudes $N^{\rm II}_3$ 
and $N^{\rm II}_4$. 
For convenience, we set
\b
N_3^{\rm II} = C_1 N_1^{\rm II} \quad ; \quad 
N_4^{\rm II} = C_2 N_1^{\rm II}  
\label{matchingmodes13}.
\e
These amplitudes $C_1$ and $C_2$ can be exactly determined by 
matching the modes and their derivatives at the singularity. 
However, since only their modulus squares are finally required, 
these can be determined from the expressions for the modulus 
square of the Bogolubov coefficients for these modes 
(see Ref.~\cite{kt99}) without doing any complicated calculation. 
\par
Now, at the boundary $u=u_2$, it is clear that positive frequency 
electric field modes go over to the positive frequency Minkowski 
modes and similarly for the negative frequency modes. 
Since the mode matching must hold for arbitrary $v$, one must 
have $\gamma_{\rm in} = \gamma_{\rm out}$.
Using this, we get
\br
N^{\rm out}_1 &=& C_1 N_1^{\rm in} \l({N_{\rm in} 
\over N_{\rm out}}\r) \, e^{i(\alpha_{\rm in}u_1 + 
\alpha_{\rm out}u_2)} 
\l(\sqrt{\smash 2 \over qE_0}\gamma_{\rm in}\r)^{-{\s 2i\alpha_{\rm in} 
\gamma_{\rm in} \over \s qE_0} + {\s 1\over \s 2}}  \nonumber \\
&& \qquad \times \; \l[ \sqrt{\smash 2 \over qE_0}\gamma_{\rm in} - 
\sqrt{\smash qE_0 \over 2}T \r]^{{\s 2i\alpha_{\rm in}\gamma_{\rm in} 
\over \s qE_0} -{\s 1\over \s 2}}
\label{matchingmodes14a}
\er
and 
\br
N^{\rm out}_2 &=& C_2 N_1^{\rm in} \l({N_{\rm in} 
\over N_{\rm out}}\r) e^{-i(\alpha_{\rm in}u_1 + 
\alpha_{\rm out}u_2)}  \l(\sqrt{\smash 2 \over qE_0}
\gamma_{\rm in}\r)^{{\s 2i\alpha_{\rm in} \gamma_{\rm in} 
\over \s qE_0} + {\s 1\over \s 2}}  \nonumber \\
&& \qquad \times \; \l[ \sqrt{\smash 2 \over qE_0}\gamma_{\rm in} - 
\sqrt{\smash qE_0 \over 2}T \r]^{-{\s 2i\alpha_{\rm in}\gamma_{\rm in} 
\over \s qE_0} -{\s 1\over \s 2}}
\label{matchingmodes14b}.
\er
From the expression for the Bogolubov coefficients, we have
\b
|C_1|^2 = 1 + e^{-\pi\lambda}  \quad ; \quad  |C_2|^2 = 
e^{-\pi\lambda},
\label{matchingmodes15}
\e
where $\lambda = (m^2 + {\bf k}_{\perp}^2)/qE_0$ as usual. 
The final expressions for the Bogolubov coefficients are
\b
\l|\alpha_{\bf k}\r|^2 = \l|{N^{\rm out}_1 
\over N^{\rm in}_1}\r|^2 
= \l({\gamma_{\rm in} \over \omega_{\rm in}}\r) 
\l({ \l(qE_0T - 2\gamma_{\rm in}\r)^2 + {\bf k}_{\perp}^2 + m^2 
\over \l(qE_0T - 2\gamma_{\rm in}\r)^2 }\r) 
\l(1 + e^{-\pi \lambda}\r) 
\label{matchingmodes16a} 
\e
and 
\b
\l|\beta_{\bf k}\r|^2 = \l|{N^{\rm out}_2 
\over N^{\rm in}_1}\r|^2 
= \l({\gamma_{\rm in} \over \omega_{\rm in}}\r) 
\l({ \l(qE_0T - 2\gamma_{\rm in}\r)^2 + {\bf k}_{\perp}^2 
+ m^2 \over \l(qE_0T - 2\gamma_{\rm in}\r)^2 }\r)\, 
e^{-\pi \lambda} 
\label{matchingmodes16b}.
\e
The expression for $|\beta_{\bf k}|^2$ gives the number of pairs 
produced for this electric field system. 
Note that the relative probability of particle production which is 
\b
{\l|\beta_{\bf k}\r|^2 \over \l|\alpha_{\bf k}\r|^2 } = 
{ e^{-\pi \lambda}\over 1 + e^{-\pi \lambda}} ,
\label{matchingmodes17}
\e 
is {\it independent} of $T$. 
Thus, in the limit of $T \to \infty$ (or equivalently 
$u_1 \to -\infty$ and $u_2 \to \infty$), which corresponds 
to an uniform electric field existing over all space and time, 
it is clear that the standard results are recovered. 
Moreover, the Bogolubov coefficients also tend to the standard 
values  upto a multiplicative factor i.e.
\b
\lim_{T \to \infty} \l|\alpha_{\bf k}\r|^2 = \l({\omega_{\rm in} - 
k_x \over \omega_{\rm in}}\r) \l(1 + e^{-\pi \lambda}\r), 
\qquad
\lim_{T \to \infty} \l|\beta_{\bf k}\r|^2 = \l({\omega_{\rm in} - 
k_x \over \omega_{\rm in}}\r) e^{-\pi \lambda}
\label{modbogolubov}.
\e
This difference can be traced to the fact that, in the standard 
calculation, boundary effects at $t\to \pm \infty$ due to mode 
matching were not taken into account. 
Therefore, the above expressions can be considered to be the 
correct ones with the extra factors arising due to these boundary 
effects at $t\to \pm \infty$. 
Note that both the coefficients have been modified by 
the same factor. 
It therefore follows that the relative probability of pair 
production, which is the ratio 
$\l(|\beta_{\bf k}|^2/|\alpha_{\bf k}|^2\r)$, is independent 
of this factor.
\par
The imaginary part of the effective Lagrangian for this system follows. 
Omitting several straightforward intermediate steps (see, for 
example Ref.~\cite{sriram97,paddyaspects}), one has
\b
\int\!d^4x {\rm Im}{\cal L}_{\rm eff} = 
{V \over 2(2\pi)^3}\sum_{n=1}^{\infty} {(-1)^{n+1} \over n} 
e^{-{\s \pi m^2 \over \s qE_0}n}
\int\! dk_x  \int\! 2\pi |{\bf k}_{\perp}|\, d|{\bf k}_{\perp}| 
e^{-{\s \pi n \over \s qE_0}{\bf k}_{\perp}^2}
\label{matchefflag1},
\e
where the limits over $k_x$ and $|{\bf k}_{\perp}|$ are to be 
determined from the condition in Eq.~(\ref{matchingmodes6}). 
From Eq.~(\ref{matchingmodes6}), we have,
\b
{\bf k}_{\perp}^2 < (qE_0T)^2 + (2qE_0T)k_x - m^2 = L(k_x).
\label{matchefflag2}
\e
Thus, the limits for the $|{\bf k}_{\perp}|$ integral are 
$\l(0, \sqrt{\smash L(k_x)}\r)$. From the condition that 
$L(k_x) > 0$ always, one must have 
\b
k_x > - \l({(qE_0T)^2 - m^2 \over 2qE_0T}\r) = - M
\label{matchefflag3}.
\e
Consider only the integral over $k_x$ and $|{\bf k}_{\perp}|$. 
Integrating over $|{\bf k}_{\perp}|$, we get
\br
I &=& \int_{-M}^{\infty}\! dk_x  \int_{0}^{\sqrt{L(k_x)}}\! 
2\pi |{\bf k}_{\perp}|\, d|{\bf k}_{\perp}| 
e^{-{\s \pi n \over \s qE_0}{\bf k}_{\perp}^2}  \nonumber \\
&=& {qE_0 \over n}\int_{-M}^{\infty}\! dk_x \l[1 - 
e^{-{\s \pi n \over \s qE_0}L(k_x)}\r]
\label{matchefflag4}.
\er
The integral over the first term in the square brackets 
gives a formally divergent term. 
Denoting this by $Z$ and doing the integration over $k_x$, 
one obtains
\b
I = {qE_0 \over n}\l\{ Z + \l({(qE_0T)^2 - m^2 \over 2qE_0T}\r) - 
\l({1 \over 2\pi n T}\r) \r\}
\label{matchefflag5}.
\e
Substituting the above into (\ref{matchefflag1}), we have
\br
\int\!d^3{\bf r}\, dt \, {\rm Im}{\cal L}_{\rm eff} 
&=& V\sum_{n=1}^{\infty} {1\over 2} {(qE_0)^2 \over (2\pi)^3} 
{(-1)^{n+1} \over n^2} \exp\l[-{\pi m^2 \over qE_0}n\r] 
\nonumber \\
&& \qquad \times \; \l\{ Z + {1 \over 2}T - {m^2 \over 2(qE_0)^2T} - 
{1 \over 2\pi qE_0n T} \r\}
\label{matchefflag6}.
\er
(Note that in the definition of the vacuum persistence amplitude, 
the limits $t_1 \to -\infty$ and $t_2 \to \infty$ were considered. 
Since two independent limits are being taken, it follows that, 
when differentiating with respect to $T$ in Eq.~(\ref{matchefflag6}), 
the right hand side has to be multiplied by a factor $1/2$.) 
Differentiating both sides with respect to $T$ and discarding 
the dimensionless divergent term $Z/T$ , one gets the result  
\b
{\rm Im}{\cal L}_{\rm eff} = \sum_{n=1}^{\infty} {1\over 2} 
{(qE_0)^2 \over (2\pi)^3} {(-1)^{n+1} \over n^2} 
e^{-{\s\pi m^2 \over \s qE_0}n} \l\{ 1 + {m^2 \over (qE_0)^2T^2} + 
{1 \over \pi qE_0n T^2} \r\}
\label{matchefflag7}.
\e
This expression for the effective Lagrangian shows boundary effects 
occurring as a correction to the standard result. 
Taking the limit of $T\to \infty$ does reproduce the standard result. 
The positive sign for these correction terms implies that the 
vacuum persistence probability is {\it smaller} than that for 
the uniform electric field case.  
\par
We conclude this section by making a few general remarks regarding 
the importance of the light cone gauge in calculating the 
effective Lagrangian. 
The use of this gauge suggests a way of calculating the effective 
Lagrangian for a constant electric field system. (By a constant 
electric field system one means any spacetime region that has a constant 
electric field present in it.) 
The following points may be deduced from the effective Lagrangian 
calculation presented in this section: 
(1)~The light cone gauge, as discussed earlier, has the property 
that the electric field modes are singular at finite spacetime 
points that are located on null lines. 
(2)~Each singularity indicates the energy of the vacuum mode that 
is excited by the electric field. Particle production occurs in that mode when 
the vacuum modes propagate past this singularity. 
The set of all these singularities determines the range of energies 
for which particles are produced (see the discussion leading to 
condition Eq.~(\ref{matchingmodes6})). 
(3)~In the calculation of the Bogolubov coefficients, the boundary 
conditions present in the system modified both the coefficients 
in exactly the same way (see Eq.~(\ref{matchingmodes16b})). 
This implies that the relative probability of pair creation 
which is the ratio $|\beta_{\bf k}|^2/|\alpha_{\bf k}|^2$ is 
{\it independent} of the extra factors introduced by the 
boundary conditions. 
It is dependent solely on the presence of the singularities. 
This point is justified for any constant electric field system 
if one accepts that particle production in an electric field is 
essentially a tunnelling process which arises, in this case, 
because of the singularities present on the light cone. 
(In any other gauge, like the time dependent gauge for example, 
the inverted oscillator nature of the effective potential is 
responsible for particle production.) 
\par
We can use these three points to determine a procedure to 
calculate the effective Lagrangian for an arbitrary constant 
electric field system in the following manner. 
Consider a region of spacetime that has a constant electric field 
of magnitude $E_0$ along, say, the $\hat{\bf x}$ direction 
(without any loss of generality). 
The field modes in this region can always be described in terms 
of the modes of the light cone gauge (multiplied by a 
suitable gauge factor). 
The only singularities that occur in the mode functions are 
due to the light cone gauge modes (we assume that singular gauge 
transformations are not allowed). 
Now, identify the set of null rays $u=constant$ passing 
through this region. Since singularities occur on these rays, 
this set determines the set of possible modes described by 
the wave vector ${\bf k}$ that can be excited by the electric field. 
That is, the possible range of values that can be taken by 
${\bf k}=(k_x,k_y,k_z)$ are determined. 
Using the fact that the relative probability of pair production 
in a particular mode, given by 
\b
{|\beta_{\bf k}|^2\over |\alpha_{\bf k}|^2} = {e^{-\pi\lambda} 
\over 1+ e^{-\pi\lambda} },
\label{matchefflag8}
\e
is independent of extra factors arising from boundary conditions, 
the effective Lagrangian can be calculated using the formula in 
Eq.~(\ref{matchefflag1}) by evaluating the integrals over ${\bf k}$ 
over the range of values determined above. Any formally divergent 
terms that occur have to be discarded.  
\par
We apply the procedure outlined above to two simple electric 
field systems. 
The first is the finite time constant electric field configuration. 
Here, the electric field is switched on over all spatial points 
for a finite time interval $T$. It is easy to see that all 
null rays $u=constant$ intersect this region. 
Hence, the range of possible values for ${\bf k}=(k_x,k_y,k_z)$ 
is from $(-\infty, \infty)$ for all the three components. 
Carrying out the integration over $k_y$ and $k_z$, we obtain
\b
\int_0^T\! dt \, {\rm Im}{\cal L}_{\rm eff} = \sum_{n=1}^{\infty} 
{1\over 2} {(qE_0)^2 \over (2\pi)^3} {(-1)^{n+1} \over n^2} 
e^{-{\s\pi m^2 \over \s qE_0}n} \int_{-\infty}^{\infty}\! 
{dk_x \over qE_0}
\label{matchefflag9}
\e
Since this expression is formally divergent, we use the 
following regularization procedure to obtain a finite result. 
We set
\b
\int_{-\infty}^{\infty} {dk_x \over qE_0} = 
Z + \int_0^T\! dt
\label{matchefflag10}
\e
where $Z$ is a formally divergent term. 
Differentiating both sides in (\ref{matchefflag9}) with respect 
to $T$ and neglecting the dimensionless divergent term $Z/T$, 
it is easy to show that one obtains the standard result. 
This result is not too surprising because to set up a constant 
electric field over all space requires an infinite 
amount of energy. 
Hence, vacuum modes of arbitrary energy are excited. 
The second system is an uniform electric field that is 
bounded along the $x$-axis (but not along the $y$ or $z$ axes) 
to a region of width $x_0$ but existing for all time. 
This system would correspond to a pair of capacitor plates 
(of infinite area) separated by a distance $x_0$. 
Here too, it is clear that all null rays pass through this region 
implying that all vacuum energy modes are excited. 
The effective Lagrangian for this sytem too is the standard result. 
Notice that in both the above examples, the actual number of 
pairs produced per mode cannot be determined easily since 
the boundary conditions are non-trivial and difficult to impose. 
More complicated examples can be studied in this fashion. 

\section{Conclusions}
\label{sec:electricdiscussion}
Summarising the analysis in this paper, we see that particle 
production in an uniform electric field has been described 
differently from the standard method. 
The lightcone gauge in Eq.~(\ref{eqn:lightconegauge}), clearly 
indicates the presence of a null surface which is responsible 
for particle production. 
This is very similar to the black hole case where again a null 
surface present in the manifold is responsible for particle 
creation. Both the black hole modes and the electric field modes 
(in the light cone gauge) possess a {\it logarithmic} 
singularity at the null surface which determines pair creation. 
However, the crucial difference between the two cases is that 
the null surface in the black hole case is a {\it one way} 
surface which is the horizon. 
To obtain particle production, we need the semi-classical 
prescription discussed in \cite{kt99} or some other equivalent 
prescription that takes into account this one-way nature 
of the horizon. 
In the electric field case, the modes can be described either 
by the gauge in Eq.~(\ref{eqn:lightconegauge}) which is written 
in terms of the ``right moving'' null coordinate $u=t-x$ 
(as done in this chapter) or by the gauge in 
Eq.~(\ref{eqn:lightconegauge1}) which uses the ``left moving'' 
null coordinate $v=t+x$ (both these gauges are appropriate 
light cone gauges). 
Whether the propagation of the electric field modes occurs 
from left to right or right to left across the singularity 
makes no difference to the final result. 
This implies that particle production in an uniform electric 
field is a genuine tunnelling phenomena (in the quantum 
mechanical sense) which is not the case for the black hole system.
\par
The effective Lagrangian, when calculated in the lightcone gauge 
(see section~(\ref{sec:electricefflag})), is found to be the same 
as the standard result.  
The modes in this gauge were also explicitly ``normalizable'' 
by a suitable physically reasonable criterion (that they reduce 
to the standard Minkowski modes in the limit of the field 
tending to zero). 
Because of this ``normalizable'' property we can look upon the 
lightcone gauge as a more {\it natural} gauge to describe the 
uniform electric field in. 
(This property also prompts the question as to whether one 
can determine indirectly a suitable ``normalization'' constant 
for the transcendental scalar field modes in the time or space 
dependent gauges.) 
One of the assumptions made in calculating the proper time kernel 
is regarding the applicability of the Feynman-Kac formula to 
such a singular system. 
However, this can be justified since, in the limit of $E_0 \to 0$, 
the free space result is obtained. 
\par
The lightcone gauge  was also used to calculate the Bogolubov 
coefficients and the effective Lagrangian for a finite time, 
but spatially non-uniform, electric field. 
The Bogolubov coefficients, in the limit of $T\to \infty$, which 
implies a uniform electric field over space and time, were found 
to reduce to the usual results upto a multiplicative factor which  
was due to extra boundary effects. 
The property of the light cone gauge that mode functions in 
this gauge are singular on null surfaces was used to develop 
a procedure to calculate the effective Lagrangian for a constant 
electric field present in an arbitrary region of spacetime. 
This procedure was based crucially on the assumption that the 
relative probability of pair creation depends only on the 
presence of the singularities and not on the specific 
boundary conditions present. 
This can be justified because pair creation in the uniform 
electric field is essentially a tunnelling process. 
In the light cone gauge, only tunnelling across the singularities 
produces particles and not otherwise. 
This procedure can be extended to arbitrary electric field 
systems which have the property that, in a sufficiently small 
neighbourhood of an arbitrary spacetime point, the electric field 
can be regarded as constant. 
This is very similar to the case of Reimannian manifolds which are 
locally flat. 
It would be therefore be of interest to ask if a suitable 
formalism can be developed on the lines of general relativity 
to describe arbitrary electric fields in the lightcone formalism 
and extend it to electromagnetic fields that satisfy the 
condition ${\bf E}^2 - {\bf B}^2 > 0$ (which must hold for 
particle production to take place). 
Such a formalism can be expected to shed light on the role of 
gauge transformations in particle production and would probably 
have relevance in describing backreaction on the 
electric field system. 
These issues will be considered in a future publication.

\end{document}